# Quantum Materials for Spin and Charge Conversion

Wei Han[1,2*], YoshiChika Otani[3,4], Sadamichi Maekawa[4,5]

[1] International Center for Quantum Materials, School of Physics, Peking University, Beijing 100871, P. R. China

[2] Collaborative Innovation Center of Quantum Matter, Beijing 100871, P. R. China

[3] Institute for Solid State Physics, University of Tokyo, Kashiwa 277-8581, Japan

[4] RIKEN Center for Emergent Matter Science (CEMS), Wako 351-0198, Japan

[5] Kavli Institute for Theoretical Sciences (KITS), University of Chinese Academy of Sciences, Beijing 100049, P. R. China

*Correspondence to: weihan@pku.edu.cn

**Spintronics aims to utilize the spin degree of freedom for information storage and computing applications. One major issue is the generation and detection of spins via spin and charge conversion. Quantum materials have recently exhibited many unique spin-dependent properties, which can be used as promising material candidates for efficient spin and charge conversion. Here, we review recent findings concerning spin and charge conversion in quantum materials, including Rashba interfaces, topological insulators, two-dimensional materials, superconductors, and non-collinear antiferromagnets. Important progress in using quantum materials for spin and charge conversion could pave the way for developing future spintronics devices.**

# Introduction to pure spin current

Spintronics aims to utilize the spin degree freedom of electrons for potential applications such as novel information storage and computing[1-3]. Compared to conventional devices using electron's charge properties, spintronic devices potentially have several major advantages, including non-volatility, faster data processing speed, higher integration densities, and less electric power consumption. The key challenges are the efficient generation and detection of pure spin current, which does not accompany any charge current, as illustrate in Fig. 1a[1]. The pure spin current could be used as information transistors and logic devices[4-6]. For example, spin-based field effect transistors can provide on and off operations via the manipulation of the spin current in the spin channel[4,7]. As shown in Fig. 1b, the pure spin current that carries the information flows from the injector to the detector via the spin channel. Due to the spin relaxation, the spin-dependent chemical potential exponentially decays during the spin diffusion (Fig. 1c). Thus, for the purpose of spin transistor and logic device applications, the spin channel with long spin lifetimes and long spin diffusion length is needed. Graphene might hold the promise for this purpose since the demonstration of its fantastic spin dependent properties at room temperature arising from its low spin orbit coupling and high mobility[8,9].

Besides serving as a transmitter of information, the spin current also carries angular momentum. As illustrated in Fig. 1a, the spin current can also be viewed as the transfer of angular momentum, which could be used for magnetization switching applications via spin transfer torque[10,11]. In the last several years, the spin orbit torque, arising from pure spin current in nonmagnetic materials (NM) (Fig. 1d), can be used to switch the magnetization of adjacent ferromagnets (FM)[12-16]. As illustrated in Fig. 1e, the spin current provides a spin transfer torque to the magnetization that can be expressed by:

$$\tau_{ST} = \frac{\hbar}{2}\hat{\mathbf{m}} \times (\hat{\boldsymbol{\sigma}} \times \hat{\mathbf{m}}) \qquad (1)$$

where $\hbar$ is the reduced Plank constant, $\hat{\mathbf{m}}$ is the magnetization vector, and $\hat{\boldsymbol{\sigma}}$ represents the spin polarization direction of the spin current. For this purpose, significant spin orbit coupling is usually required for the nonmagnetic materials, which is essential for the conversion of charge current to spin current. There are two major mechanisms for the conversion of charge to spin current, namely spin Hall effect[17-25] and Edelstein effect[26]. The major difference between these two effects is that spin Hall effect is usually considered as a bulk effect, which means that the spins diffuse in the direction perpendicular to the charge current, whereas the Edelstein effect is often referred as an interface effect that the spins and electrons are confined in the two-dimensional (2D) state. The spin Hall effect in conventional metals and semiconductors has already been described in several review articles[27,28].

In this review article, we focus on the spin and charge conversion in recent quantum materials, including spin-momentum locked Rashba interfaces and topological insulator surface states, 2D materials, quasiparticles in superconductors, as well as non-collinear antiferromagnets (Fig. 2). Due to their unique spin dependent properties, these quantum materials have exhibited interesting spin and charge conversion phenomena, such as giant spin and charge conversion efficiency, and gate tunable spin and charge conversion, etc. This article reviews those most important findings of quantum materials for spin and charge conversion, and is organized in the following order. The first section discusses the spin and charge conversion of Rashba states via Edelstein and inverse Edelstein effects at non-magnetic metal interfaces, insulating oxide interfaces, and metal-insulating oxide heterostructures. The second section discusses the spin-momentum locked surface states of $Bi_2Se_3$-type topological insulators and new types of topological insulators, including α-Sn and topological Kondo insulator SmB$_6$. The third section discusses the spin and charge conversion in 2D materials, including graphene-based heterostructures with large spin orbit coupling induced via proximity effect, and transition metal dichalcogenides (TMDCs), such as MoS$_2$ and WTe$_2$. The fourth section discusses the quasiparticle-mediated spin Hall effect in s-wave superconducting NbN thin films. The fifth section discusses the large anomalous Hall effect and spin Hall effect in non-collinear antiferromagnets. The final section discusses the perspectives of quantum materials for future spintronics applications.

**Spin and charge conversion in Rashba interfaces**

The spin and charge conversion in Rashba interfaces is often referred as Edelstein effect or Rashba-Edelstein effect; a nonzero spin accumulation is accompanied with a charge current flowing in 2D asymmetric systems[26]. At the interface, the coupling between the momentum and spin polarization directions can be expressed by the following Hamiltonian[26,29]:

$$H_R = \alpha_R(\hat{\boldsymbol{k}} \times \hat{\boldsymbol{z}}) \cdot \hat{\boldsymbol{\sigma}} \qquad (2)$$

where $\alpha_R$ is the Rashba coefficient that is proportional to the spin orbit coupling ($\lambda_{SO}$) and the electrical field ($E$) perpendicular to the interface ($\alpha_R = e\lambda_{SO}E$, $e$ is the electron charge), $\hat{\boldsymbol{z}}$ is the unit vector perpendicular to the interface, $\hat{\boldsymbol{k}}$ is the wave vector, $\hat{\sigma}$ is the spin Pauli matrices vector. As a result of interaction between spins and the momentum directions of the carriers, the energy dispersion exhibits two contours at the Fermi surface, as illustrated in Fig. 3a. These energy dispersion curves can be experimentally revealed by angle-resolved photoemission spectroscopy (ARPES). The splitting in energy dispersion curves for spin-up and spin-down carriers could be used to extract the Rashba coefficient at the interface/surface, i.e., large Rashba coefficient has been identified at the interface of Bi with other metallic materials[30,31].

The spin and charge conversion can be explained based on the shift of the Fermi contours in the presence of electric fields. For the two Fermi contours (Fig. 3a), the inner one has spin-momentum locking feature with a clockwise spin texture, while the outer one has a counterclockwise spin texture. These two Fermi contours are symmetric without an electric field. However, in the presence of an electric field, as illustrated in Fig. 3b, both of these Fermi circles shift towards positive $k_x$ direction. Shift of the outer Fermi circle gives rise to an increase in spin-up density. While shift in inner Fermi circle causes an increase in spin-down density. Usually, the net effect is a spin-up density accompanying the charge current, since the outer Fermi circle plays the major role. Quantitatively, the generated spin density ($\hat{S}$) is proportional to charge current density ($\hat{J}_C$), described by the following relationship:

$$\hat{S} \propto (\alpha_R/e\hbar)(\hat{z} \times \hat{J}_C) \quad (3)$$

Thus, for the efficient charge to spin conversion, a large Rashba coefficient is favored. Similar to the inverse spin Hall effect, the inverse Edelstein effect refers to that spin accumulation in inversion 2D asymmetric systems could generate an in-plane electric field perpendicular to the spin polarization direction for both systems with weak and strong spin orbit couplings[32,33]. As illustrated in Fig. 3c, the injection of spin-up polarized carriers results in the shift of the outer/inner Fermi circles to the positive/negative $k_x$ direction, generating a net charge current along the positive $k_x$ direction. In the last several years, intensive investigations of the Edelstein effect and inverse Edelstein effect have been performed on the Rashba-split states at the non-magnetic metal interfaces[33-36], insulating oxide interfaces[37-41], and metal-insulating oxide interfaces[42,43].

We first discuss the Edelstein effect and inverse Edelstein effect of the Rashba-split interfaces between two metallic layers. To probe the Edelstein effect at the Ag/Bi interface, Zhang *et al*[34] have used the spin-polarized positron beam to detect the spin polarization of the outermost surface electrons in the presence of a charge current. This method is ultra-sensitive to the surface spin polarization since the bound state of a positron and an electron is a Positronium. Since Positronium can only be formed for low electron density, only the outermost surface in a metal could satisfy this condition, which makes this technique ultra-sensitive to the surface. During the measurement, as illustrated in Fig. 3d, a transversely spin-polarized positron beam is generated by a $^{22}$Na source and an electrostatic beam apparatus, and then guided to inject into the Ag/Bi samples. A high-purity Ge detector is used to detect the Positronium annihilation γ rays, from which the component of the surface spin polarization can be obtained in the presence of a charge current flowing in the Rashba interfaces of Ag/Bi. Due to the large difference of the resistivity between Bi and Ag, the charge current mainly flows in the Ag layers. By systematically varying the thickness of Bi in the bilayer structures of α-Al$_2$O$_3$ (0001)/Ag (25 nm)/Bi, the charge-to-spin conversion effect of the Rahsba interface is investigated. For pure Ag thin films, the surface spin polarization is estimated to be 0.5% with $J_C$ = 15 A/m, which is mall and most likely arising from the small spin Hall effect of Ag. A giant enhancement of the surface spin polarization is probed when the Bi thickness is ~ 0.3 nm with a value of 4.1% with $J_C$ = 15 A/m, which is more than ten times higher than that of pure Ag films, indicating the dominant role of the Ag/Bi interface. Fig. 3e shows the Bi thickness dependence of surface spin polarization for the α-Al$_2$O$_3$ (0001)/Ag (25 nm)/Bi bilayer samples. Clearly, there is a strong enhancement when the Bi reaches 0.3 nm, which is approximately monolayer considering the Bi atomic radius of 0.15 nm. As the Bi thickness further increases, the detected surface spin polarization decreases, which is expected since the interface is the source for the spin generation. When the surface is far away from the interface, there is a spin depolarization due to spin scattering, following an exponential relationship ($\exp(-d/\lambda_{sd})$), where $\lambda_{sd}$ is the spin diffusion length of Bi. Based on the exponential relationship (solid line in Fig. 3e), the spin diffusion length of Bi is estimated to be ~ 2.1 nm. Since the sign of the Rashba coefficient depends on the stacking order of Bi and Ag, reversing the Bi and Ag layer is expected to give rise to a negative surface spin polarization. Consistent with this expectation, experimental results exhibit a negative sign for the α-Al$_2$O$_3$ (0001)/Bi (8 nm)/Ag (25, 100, 200, 300, 400, 500 nm) bilayer samples. A large surface spin polarization of ~ - 5% is probed when the Ag thickness is 25 nm with $J_C$ = 15 A/m. As the Ag layer thickness increases, the surface spin polarization decreases exponentially due to the spin relaxation in Ag. Since the spin orbit coupling in Ag is much weaker than that of Bi, a much longer diffusion length of ~357 nm is observed for the surface spin polarization of the α-Al$_2$O$_3$ (0001)/Bi (8 nm)/Ag bilayer samples. The opposite spin polarization and its dependence on the thicknesses of Bi and Ag layers have demonstrated the dominant role of the Rashba interfaces between Bi and Ag in charge-to-spin conversion results.

The inverse Edelstein effect, spin to charge conversion, of Ag/Bi Rashba interface states has been investigated by the Rojas-Sanchez *et al*[33]. To generate the spin current in the Rashba interfaces, a thin NiFe (Py) layer is grown on the Bi/Ag, which acts as a spin pumping source. The spins are generated from the Py layer under its magnetization resonance conditions, which results in the generation of spin currents at the interface. This dynamic spin injection method is called spin pumping, which is a well-established technique to inject spin polarized carriers into various materials, including metals, semiconductors, Rashba interface states, etc[23,25,33,44,45]. It can be considered as a process of transferring angular momentum from the FM layer into the adjacent NM layer, and the angular momentum in the NM layer is mediated by pure spin current. Fig. 3f shows the schematic of the measurement setup for probing the inverse Edelstein effect in the Bi/Ag bilayer samples. At the resonance magnetic fields for the Py magnetization under certain radio frequencies, a DC pure spin current is injected into Ag and arrives at the Bi/Ag Rashba interface states; one type of spin polarized carriers move towards the Py, and the opposite type of spin polarized carriers move away from the Py. At the Bi/Ag Rashba interface states, the pure spin current is converted into a charge current in the sample's plane ($I_C$) due to the inverse Edelstein effect, which is probed by a voltage meter. The Bi/Ag samples are grown on a Si/SiO$_2$ wafer substrate (~ 500 nm thermally oxidized SiO$_2$) by successive evaporation of Bi, Ag and NiFe in ultra-high vacuum. The texture of the Bi layer is along the crystal's (111) direction in the rhombohedral notation, which is favored for the spin to charge conversion since relatively large Rashba coefficients have been observed by ARPES at the interface for (111)-oriented Bi thin films and their heterostructures[30,31]. Fig. 3g shows the ferromagnetic resonance (FMR) spectra and measured charge currents as a function of the external magnetic field for three typical samples: 10 nm Ag/15 nm Py (left), 8 nm Bi/15 nm Py (middle) and 8 nm Bi/5 nm Ag/15 nm Py (right). For the similar FMR absorption signal, the measured charge currents of these three samples exhibit large variations. For pure Ag, the measured charge current is negligible (left), consistent with the small spin Hall angle in Ag films. For Bi films with a relatively larger spin Hall angle compared to Ag, a larger charge current is observed (middle). The largest value of the charge current is obtained on the 8 nm Bi/5 nm Ag/15 nm Py sample. Given the fact that injected spin current densities are similar for all the samples indicated by similar enhanced Gilbert damping of Py layer, the observation of largest charge current on the Bi/Ag Rashba interface strongly supports the significant spin to charge conversion arising from the large Rashba coefficient of the Bi/Ag Rasbha interface states. Besides the Bi/Ag interface, the spin and charge conversion of other Rashba interfaces, including Sb/Ag and Bi/Cu, has also been investigated via either spin pumping method or spin absorption method in a Cu-based nonlocal spin valve[35,36].

Since the pioneering work that reported the high mobility conducting two-dimensional electron gas (2DEG) between two insulating oxides (SrTiO$_3$ and LaAlO$_3$)[46], many interesting physical properties have been discovered[47,48]. A particular interesting point is that all these physical properties are gate-tunable via a perpendicular electric field. This oxide interface becomes conducting when the LaAlO$_3$ layer exceeds a critical thickness of 3-4 unit cells. As illustrated in Fig. 4a-b, the polar oxide LaAlO$_3$ consists of two layers, (LaO)$^+$, and (AlO$_2$)$^-$, and the top surface of SrTiO$_3$, a thin TiO$_2$ layer, becomes conducting. To account for the high mobility 2DEG at the interface, several mechanisms have been proposed, including the "polar catastrophe", an electric field with thickness of the polar LaAlO$_3$ layer resulting in charge transfer from the LaAlO$_3$ surface to the interface[49], cation intermixing giving La doping in the SrTiO$_3$ surface[50], and oxygen off-stoichiometry to form SrTiO$_{3-x}$ conducting layer[51]. Particularly intriguing for spintronics, a large Rashba coefficient has been reported via weak anti-localization measurements at $T = 1.5$ K[52] and it can be markedly modulated from $1 \times 10^{-12}$ to $5 \times 10^{-12}$ eVm by applying an electric field. Recently, the inverse Edelstein effect of the oxide interface has been achieved experimentally and its spin and charge conversion efficiency is highly gate tunable[37,38].

Song *et al*[38] performed the room temperature inverse Edelstein effect experiments on the SrTiO$_3$ and LaAlO$_3$ interface using a thin Py metallic layer (~ 20 nm) as the spin pumping source. The spins are injected via spin pumping into the oxide interface, and are converted to a charge current that could be measured using a voltage meter. The angular momentum is transferred from Py to the 2DEG across the LaAlO$_3$ layer via spin tunneling across the LaAlO$_3$ layer and/or localized states in the LaAlO$_3$ layer (i.e. oxygen vacancies). To achieve the gate tunable inverse Edelstein effect, a sample of the critical thick LaAlO$_3$ layer of 3-unit cell is used, of which the interfacial conductivity can be modulated dramatically by an electric field at room temperature. During the spin pumping measurement, an electrode of silver paste is used on the other side of SrTiO$_3$ substrate to serve as a back gate, as schematically shown in the inset of Fig. 4c. The gate tunable spin signal of the Rashba 2DEG between SrTiO$_3$ and 3-unit cell LaAlO$_3$ has been observed (Fig. 4c). It is clearly seen that the inverse Edelstein effect voltage ($V_{IEE}$), resulting from spin to charge conversion at the Rashba-split 2DEG, can be tuned markedly by the gate voltage. Under negative gate voltages, the $V_{IEE}$ is significantly lower, which means a lower effective spin-to-charge conversion. Whereas, the large spin signal and low resistance of the 2DEG under positive gate voltage indicate a larger spin-to-charge conversion. This observation is attributed to the gate tunability of the carrier density, Rashba coefficient at the interface, and the spin pumping efficiency that is related to the spin mixing conductance between Py and the Rashba 2DEG. Furthermore, the gate dependence of the spin to charge conversion has been theoretically investigated[39], which shows a good agreement with the experiential results at room temperature.

Since the dielectric constant of SrTiO$_3$ increases significantly at low temperature, the gate voltage modulation of the $V_{IEE}$ at low temperatures is expected to be significantly enhanced.

Indeed, at low temperature (T = 7 K), Lesne et al[37] reported a large gate tunable inverse Edelstein effect for the Rahsba-split 2DEG between $SrTiO_3$ and 2-unit cell $LaAlO_3$, where the spin to charge conversion sign could be even reversed. The 2-unit cell $LaAlO_3$ is able to produce a conducting Rashba 2DEG because the critical thickness of $LaAlO_3$ for the formation of the conducting 2DEG is reduced due to the deposition of ferromagnetic metallic layers, including both Co and Py[37,53]. A large spin to charge conversion efficiency ($\lambda_{IEE} = J_C^{2D}/J_S^{3D}$) of 6.4 nm is reported, which is more than one order larger than that of Rashba interface states between metallic layers[33]. The gate dependence of the spin to charge conversion efficiency is shown in Fig. 4d, which clearly demonstrates the dramatic modulation of the spin to charge conversion via the perpendicular electric field. $\lambda_{IEE}$ first exhibits little variation at negative gate voltages, and then changes the sign when the gate voltage is positive. The maximum value for $\lambda_{IEE}$ is observed at the gate voltage of +125 V. This non-trivial gate dependence is attributed to the multiband property of the electronic structure for the Rashba 2DEG at the oxide interface. As the gate voltage modulates the carrier densities, the Fermi level shifts from a single band with $d_{xy}$ character to other d bands, such as $d_{xy}$ and $d_{yz}$. This experimental observation is supported by first principle theoretical studies that calculate a weak Rashba coefficient with a negative sign for the $d_{xy}$ band, and a much larger Rashba coefficient for the $d_{xy}$ and $d_{yz}$ bands with positive sign. The carrier density across Lifshitz points corresponds to ~ $1.8 \times 10^{13}$ cm$^{-2}$, as indicated by the schematic of the band structure for the oxide interface (inset of Fig. 4d). To further confirm the scenario above, the detailed band structures are essential, which can be revealed by spin-ARPES measurements.

For potential applications such as magnetic switching via spin orbit torque, the Edelstein effect, charge to spin conversion, has to be demonstrated. Recently, a significantly large spin orbit torque arising from the Edelstein effect at the SrTiO$_3$ and LaAlO$_3$ interface on an adjacent CoFeB layer is reported at room temperature[41]. Beyond the spin and charge conversion, the oxide-based interface and materials can also serve as a channel for the conducting spins, and the spin injection has been successfully performed[54-56].

Besides the Rashba states for metal-metal and oxides interfaces, the interfaces between metal and oxide have also been investigated for spin and charge conversion[42]. Fig. 4e shows the measurement geometry and the sample structure, where Py layer is the ferromagnetic layer for spin pumping source, and the Cu/Bi$_2$O$_3$ is the Rashba interface that converts spin current into in-plane charge current via inverse Edelstein effect. As shown in Fig. 4f, a clear spin pumping voltage owing to the spin-to-charge current conversion is observed around the resonance magnetic field (red curve). Whereas, no spin pumping voltages could be observed on the control samples of Py/Cu (black curve) and Py/Bi$_2$O$_3$ bilayers. This is understandable because Bi$_2$O$_3$ is an insulator and Cu does not contribute to the conversion due to its low spin orbit coupling and weak spin Hall effect. Hence, the observed spin signal on Py/Cu/Bi$_2$O$_3$ sample can only be explained by the inverse Edelstein effect of the Rashba interfacial states. A spin to charge conversion efficiency ($\lambda_{IEE}$) of ~0.6 nm is obtained, about twice the value of Bi/Ag Rashba interface[33]. Then the interfacial Rashba coefficient is determined to be − (0.46 ± 0.06) eV·Å, which might be controlled by an electric field effect similar to gate tunable inverse Edelstein effect observed at the oxide interface[37,38]. This Rashba interface can also lead to a new type of magnetoresistance, namely Edelstein magnetoresistance[43], which has similar origins as spin Hall magnetoresistance observed at the interface between FM and NM with large spin Hall effect[57,58]. This experimental work also reports that about only two thirds of the total spin relaxation occurs at the interface[43], which points an interesting direction for future work towards the 100% contribution from pure interface states.

**Spin and charge conversion in topological insulators**

Topological insulators, a class of quantum materials, have protected conducting surface or edge states, possibly arising from strong spin-orbit coupling and time-reversal symmetry[59,60]. The 2D topological insulators are quantum spin Hall insulators with the transport of spin and charge only in gapless edge states. Similar to 2D topological insulators, three-dimensional (3D) topological insulators have insulating bulk states and conducting surface states with spin polarization direction perpendicular to the momentum direction. This feature of the surface states is called spin-momentum locking, and has been first demonstrated by scanning tunneling microscopy and spin-ARPES[61,62]. Ferromagnetic tunneling contacts have also been used to detect the spin current generation due to the spin-momentum locking property of the topological surface states[63-67]. Since the spin-dependent chemical potentials are different between spin-up and spin-down carriers, the spin dependent voltage measured between the ferromagnetic electrodes and the surface states is dependent on whether the magnetization is parallel or antiparallel to the spin polarization direction[68].

In the context of spintronics, this spin-momentum locking property could be very useful for the magnetization switching via spin orbit torque[69-72]. Since the spin and the momentum directions are strongly coupled to each other in the surface states of 3D topological insulators[73-75], a spin and charge conversion efficiency of 100% could be realized in ideal case. Fig. 5a shows the schematic of the energy dispersion of the topological surface states. For both electron and hole carriers, the spin polarization directions are perpendicular to the momentum directions but with opposite spin textures. The spin texture direction is the property of a material that depends on the sign of surface/interfacial Rashba parameters[76]. For $Bi_2Se_3$-based topological insulators, the electron band has a clockwise spin texture, which results in a spin-down polarization coupled to a positive momentum direction, as illustrated in Fig. 5b. While the hole band has a counterclockwise spin texture, which results in a spin-up polarization coupled to a positive momentum direction, as illustrated in Fig. 5c. Due to the presence of a charge current, the Fermi circle shifts, giving rise to the accumulation of spin-polarized carriers, which is similar to the Edelstein effect of Rashba interfaces. For *n*-type carriers (electrons), on applying an electric field along negative $k_x$ direction, the Fermi circle shifts to the positive $k_x$ direction (from dashed to solid circles) and spin-down electrons are accumulated at the Fermi surface, as shown in Fig. 5b. For *p*-type carriers (holes), on applying an electric field along negative $k_x$ direction, the Fermi circle shifts to the positive $k_x$ direction (from dashed to solid circles) and spin-up electrons are accumulated at the Fermi surface, as shown in Fig. 5b. The spin accumulation at the surface of topological insulator ($\langle \delta S_0 \rangle$) can be expressed by[72]:

$$\langle \delta S_0 \rangle = \frac{\hbar}{2} k_F \delta k_x = \frac{\mu k_F^2 \hbar E_x}{2 v_F} \qquad (4)$$

where $k_F$ is the Fermi wavenumber, $\delta k_x$ is the shift of Fermi circle, $\mu$ is the mobility of the topological surface states, $E_x$ is the applied electrical field, and $v_F$ is the Fermi velocity. Due to the opposite sign of charge for electrons and holes, the spin and charge conversion efficiency is expected to have the same sign.

The Edelstein effect of the topological insulator surface states has been investigated using spin-torque ferromagnetic resonance (ST-FMR) technique[69,71,72]. The main idea of the ST-FMR technique is based on the ratio of spin-orbit torque due to the spin current and the Oersted field torque arising from the flow of the charge current, hence, the charge to spin conversion efficiency ($q_{ICS} = J_S^{3D}/J_C^{2D}$) could be obtained. A pioneering work, done by Mellnik et al[69], has demonstrated a giant charge to spin conversion efficiency in the $Bi_2Se_3$/Py bilayer structure at room temperature, which is much larger than the value observed in heavy metallic systems, including Pt, β-Ta, β-W, and Bi-doped Cu. Consequently, the Fermi level dependence of the topological insulators $(Bi_{1-x}Sb_x)_2Te_3$ is studied to investigate the contributions from carriers in the surface states and bulk states[72]. Fig. 5d illustrates the ST-FMR measurement of the spin current generated from the topological insulators in the $(Bi_{1-x}Sb_x)_2Te_3$ (8 nm)/Cu (8 nm)/Py (10 nm) trilayer structure, where the Cu spacing layer provides protection of the topological surface states from degradation during the deposition of the Py directly onto it. The generated spin current, carrying angular momentum, flows into the Py layer, and exerts a spin-transfer torque on the magnetization of Py controlled by an external magnetic field with an angle (θ) from the flow of the charge current ($I_C$). Fig. 5e shows the Fermi level dependence of $q_{ICS}$ for $(Bi_{1-x}Sb_x)_2Te_3$, where the position of the Fermi level is controlled by the doping ratio of Sb over Bi[77]. In the bulk insulating $(Bi_{1-x}Sb_x)_2Te_3$ samples (x from 0.5 to 0.7) at low temperature (T = 10 K), the charge-to-spin conversion effect via the Dirac surface states is almost constant, which is consistent with the theoretical expectation that the $q_{ICS}$ is inversely proportional to the Fermi velocity. For the bulk conducting $(Bi_{1-x}Sb_x)_2Te_3$ samples (x = 0 or 1), the estimated values of $q_{ICS}$, using the same analytical method, are found to be roughly equal to, or larger than those for the bulk insulating films, which might be overestimated. When the Fermi level approaches the Dirac point, $q_{ICS}$ is remarkably reduced, which might be related to the inhomogeneity of Fermi wavenumber and/or instability of the helical spin structure. The conversion from spin current to charge current, inverse Edelstein effect, of the $Bi_2Se_3$-based topological insulators has been studied via spin pumping from either ferromagnetic metallic layer or ferromagnetic insulators[78-83].

Beyond $Bi_2Se_3$-based topological insulators, recent theoretical and experimental studies have identified new topological insulators, such as α-Sn and strongly correlated Kondo insulator $SmB_6$. For α-Sn thin films, theoretical calculations show that α-Sn thin films could be topological insulators via strain or quantum-size effects[84,85]. Consequently, the helical spin polarization of the Dirac-cone-like surface states of (001)-oriented single crystalline α-Sn thin films, grown on InSb(001) substrates by molecular beam epitaxy, has been observed by ARPES measurements (Fig. 6a), which demonstrated that (001)-oriented α-Sn thin film is a new 3D topological insulator[86]. To study the spin and charge conversion in α-Sn topological insulator, Rojas-Sánchez et al[87] performed spin pumping experiments on the (001) α-Sn/Ag/Fe sample, where the Ag layer (2 nm) is used to protect the topological surface states of α-Sn from the damage caused by growing Fe directly onto it. Fig. 6b shows the FMR spectrum of the Fe and converted charge current measured on (001) α-Sn (30 monolayers)/Ag (2 nm)/Fe (3 nm)/Au (2 nm) sample. The spin to charge conversion efficiency is estimated to be ~ 2.1 nm, which is significantly larger than Rashba interfaces, such as Ag/Bi[33].

$SmB_6$, a Kondo insulator, has been theoretically proposed to be a topological insulator at low temperature due to the band inversion between 4f and 5d orbitals[88,89]. Since then, several experimental studies have demonstrated that $SmB_6$ is a new type of topological insulator, including spin-APERS technique, Hall transport methods, quantum oscillation measurements, etc[90-95]. A unique property of $SmB_6$ is that below ~ 3 K, the bulk states are insulating, and only surface carriers contribute to the conduction, as illustrated in Fig. 6c, which can be evident by the surface Hall measurement and saturation of the resistance of bulk $SmB_6$ single crystals (Fig. 6d).

Hence, SmB$_6$ provides a platform for the clear demonstration of the Edelstein/inverse Edelstein effects for the spin-momentum locked surface states, which is quite challenging for Bi$_2$Se$_3$-based topological insulators due to the presence of unavoidable bulk carriers. Song et al[96] observed the spin to charge conversion of the pure surface states in SmB$_6$ where the spin current is generated via the spin pumping from Py, as illustrated in Fig. 6d inset. The temperature dependence of the spin pumping and inverse Edelstein effect is performed to investigate how the spin voltage evolves as the surface states emerge and become dominant. As shown in Fig. 6e, the spin voltage steadily decreases as the temperature increases, and when the temperature reaches 10 K, no obvious voltage could be detected. This observation is expected since pure surface states exist below ~ 3 K, giving rise to an in-plane electrical voltage that is perpendicular to the spin directions, due to the inverse Edelstein effect. While more bulk carriers are activated as the temperature increases, the spin voltage is greatly suppressed due to the current shunting effect of the bulk SmB$_6$ crystal. Compared to Bi$_2$Se$_3$-based topological insulators, both α-Sn and Kondo insulator SmB$_6$ have counterclockwise helical spin configuration of the electron-type Fermi contours, resulting in the opposite sign of spin to charge conversion signal[76,86,87,93,96].

Up to date, a large range of the spin and charge efficiency values from ~ 0.001 to ~ 2 has been reported for the topological insulators, using different techniques including ST-FMR, spin pumping and inverse Edelstein effect, magnetization switching, and second harmonic measurement, etc[66,69-72,78-83,87,97,98]. Even for the same measurement on the same topological insulator, ie, second harmonic measurement of the spin-orbit torque arising from 3D topological insulators, different values have been reported[70,99]. One possible reason is related to the multiple sources of second harmonic voltages, i.e., large unidirectional magnetoresistance is also reported at the topological insulator/FM interface[100].

To fully address the disagreement of the exact value for spin and charge conversion efficiency generated from the surface states of 3D topological insulators, future experimental and theoretical studies are essential. Despite this disagreement, it has been demonstrated that 3D topological insulators can be used to perform efficient magnetization switching for FM films with both in-plane and perpendicular magnetic anisotropies at room temperature[97,98].

**Spin and charge conversion in 2D materials**

The isolation of graphene and related 2D materials has created new opportunities for spintronics. For example, graphene is potentially useful for spin channels in spin transistor and logic devices due to long spin lifetime and spin diffusion length[8,9,73,101]. TMDCs are potentially useful for spintronics and valleytronics devices arising from the spin-valley coupling[102,103]. Regarding the spin current generation/detection, it has been recently demonstrated that 2D materials can be potentially used for efficient spin and charge conversion, including graphene with large spin orbit coupling due to proximity effect with other materials, TMDCs with low crystalline symmetry ($MoS_2$ and $WTe_2$).

Although the intrinsic spin orbit coupling is rather weak in pristine graphene, it can be largely enhanced via proximity effect with other materials that have large spin orbit coupling, such as ferromagnetic insulator, Yttrium Iron Garnet (YIG)[104-106], and TMDCs[107,108], etc. Recently, it has been shown that YIG/graphene could be a promising material candidate for efficient spin and charge conversion[105,106]. The spin current is generated via spin pumping from YIG thin film grown on GGG substrates (Fig. 7a), and then converted to a charge current in graphene due to the enhanced spin orbit coupling and inverse Edelstein effect. The spin signal arising from the charge current flowing in graphene is measured via two Ag contacts at the ends of graphene channel. The typical magnetic field dependence curves of the spin pumping voltage measured on YIG/Graphene are shown in Fig. 7b. When the external magnetic field is applied along the graphene channel ($\phi = 90°$), the pumped spin polarized carriers will be converted to a charge current perpendicular to the graphene channel. Thus, no spin pumping voltage could be detected (black line in Fig. 7b). When the external magnetic field is applied perpendicular to the graphene channel ($\phi = 0°$ and $180°$), the pumped spin polarized carriers will be converted to a charge current with a direction along the graphene channel, giving rise to spin pumping voltages (red and blue lines in Fig. 7b). Furthermore, ionic liquid gating provides a large electric field on the graphene surface, which has been demonstrated to strongly modulate the spin to charge conversion efficiency of YIG/graphene[106].

Different from pristine graphene, TMDCs are usually with large spin orbit coupling arising from the large intrinsic spin orbit coupling of the heavy elements, such as Mo and W[102]. As a result, the spin splitting at the surface due to broken inversion symmetry could be used for converting spin current to/from charge current and switching the magnetization of nanomagnets. Both ST-FMR technique and spin pumping method have been used to study the spin and charge conversion in monolayer $MoS_2$[109,110]. The ST-FMR results on monolayer $MoS_2$/Py exhibit a large spin orbit torque generated by the Edelstein effect of thin $MoS_2$ films. Specifically, a very large damping-like torque, relative to a field-like torque, has been identified[109]. The spin pumping results on monolayer $MoS_2$/Al (3 nm)/Co (10 nm) present a large spin to charge conversion efficiency that corresponds to an effective spin Hall angle of ~ 12. Besides, the spin to charge conversion efficiency could be manipulated via a back gate voltage across a 300-nm insulating $SiO_2$ layer[110].

Very interestingly, recent experimental results show a large out-of-plane spin orbit torque in WTe$_2$ arising from the low crystalline symmetry[111]. The surface crystalline structure of WTe$_2$ belongs to the space group P$_{mn2_1}$, and the crystal structure near the surface is shown in Fig. 7c. For the WTe$_2$/Py bilayer samples, a mirror symmetry exists with respect to the *bc* plane (dashed red line), but does not exist with respect to *ac* plane, which means that there is no 180° rotational symmetry about the *c*-axis (perpendicular to the sample plane). As a result, an out-of-plane spin orbit torque is allowed by symmetry for the current flowing along the *a*-axis of a WTe$_2$/Py bilayer. Specifically, a charge current flowing along positive *a*-axis gives rise to an extra spin orbit torque that is along negative *c*-axis, while a charge current flowing along negative *a*-axis gives rise to an extra spin orbit torque that is along positive *c*-axis. As the in-plane magnetic-field angle ($\phi$) changes with current applied parallel to the a-axis, the symmetric voltage arising from in-plane spin orbit torque has a two-fold rotational symmetry, which changes sign when the magnetization is rotated in-plane by 180°. This is expected since the in-plane spin orbit torque is proportional to the angle between magnetization and in-plane spin polarization due to the charge current via spin Hall effect and/or Edelstein effect. However, the magnetic field angle dependence of the asymmetric voltage due to out-of-plane spin orbit torque does not have a two-fold rotational symmetry, thus cannot be well fitted by the $\cos(\phi)\sin(2\phi)$ relationship expected for pure Oersted field torque, as shown in Fig. 7d. This result indicates the existence of another out-of-plane spin orbit torque besides the Oersted field torque. This extra out-of-plane torque changes sign when the charge current switches direction, and is allowed for WTe$_2$/Py bilayer devices which also do not have two-fold symmetry when the current is along *a*-axis. If the charge current flows along *b*-axis WTe$_2$/Py bilayer devices, a two-fold symmetry of the asymmetry voltage is observed. These results demonstrate the possibility of generating efficient spin orbit torque based on the TMDCs and other materials with broken crystal symmetry and significant spin orbit coupling.

**Quasiparticle-mediated spin Hall effect in superconductors**

Superconductor spintronics is an emerging field focusing on the interplay between superconductivity and spintronics[112]. In the FM/superconductor heterostructures, the magnetization of FM layer can strongly modulate the superconducting critical temperature of the superconducting layer[113-116]. Spin dynamics of superconducting films could be measured via spin pumping from an adjacent ferromagnetic insulating layer, which has been recently theoretically proposed and experimentally investigated[117,118]. The underlying mechanism is that the magnetization dynamics, probed by FMR and spin pumping, is coupled to the spin dynamics of the superconductor via the interfacial s-d exchange interaction. In the content of spintronics, the spin polarization of a FM material can be probed using superconducting point contact via the differential conductance measurement since the Andreev reflection at the interface is determined by the spin polarized density of states[119]. Recently, many extraordinary spin-dependent physical properties of superconductors have been observed. For instance, large magnetoresistance has been achieved in the spin valve consisting of a superconducting layer between two ferromagnetic films[115]. This large magnetoresistance is attributed to the different superconducting temperatures for the parallel and anti-parallel states of the two FM layers' magnetization. Besides, extremely long spin lifetimes and large spin Hall effect have been discovered for superconductors mediated by quasiparticles, the superpositions of electron-like and hole-like excitations. The values of the spin lifetime and spin Hall effect exceed several orders compared to those of the normal states above the superconducting temperature[120-122]. Here, we discuss experimental results of the quasiparticles-mediated large spin Hall effect in s-wave superconducting NbN thin films.

Quasiparticle-mediated spin Hall effect of a s-wave superconductor is first theoretically studied, which predicts a large enhancement below the superconducting critical temperature[120,123]. Experimentally, Wakamura et al[122] have demonstrated the large spin Hall effect of quasiparticles in NbN thin films, using the spin absorption technique based on a Cu nonlocal spin valve[24,124]. As illustrated in Fig. 8a, the spin current is injected from Py electrode (spin injector), and then diffuses inside the Cu spin channel. The spin polarization direction is controlled via the external magnetic field, which sets the magnetization direction of the Py electrode (spin injector). When the spin current arrives at the Cu/NbN interface, spins are absorbed by the NbN with significant spin orbit coupling. Different from Cu, where spins are carried by electrons, the spins in NbN are carried by quasiparticles, and converted to a quasiparticle-mediated charge imbalance ($J_Q$) due to the inverse spin Hall effect. The spin absorption at the Cu/NbN interface can be viewed as the angular momentum transfer between the spins of electrons in the Cu and the spins of quasiparticles in superconducting NbN. The inverse spin Hall signal has been investigated across the superconducting critical temperature of NbN. Above the superconducting temperature ($T$ = 20 K and higher), a regular inverse spin Hall effect is observed and the spin Hall angle is estimated to be ~ 0.9%. The spin Hall angle exhibits tiny difference between 20 K and 300 K (~ 0.9% to ~1.3%) and no dependence on the charge current is observed. However, below the superconducting critical temperature, a large inverse spin Hall signal is observed at $T$ = 3 K, which is more than 2,000 times larger compared to the inverse spin Hall signal at 20 K. Fig. 8b compares the current dependences of the inverse spin Hall signal between $T$ = 3 K and 20 K, where the inverse spin Hall resistance is normalized by the value at 20 K ($\Delta R_{ISHE}^{normal}$) under the spin injection current of 300 μA. The strong spin current dependence of inverse spin Hall signal for quasiparticles is attributed to the increase of the local temperature at the Cu/NbN interface due to the Joule heating. Hence, when a large spin injection current is applied, the real temperature around the Cu/NbN interface increases, which results in the large suppression of the superconducting gap of NbN at the interface.

A critical parameter for the quasiparticle-mediated spin Hall effect is the length scale ($\lambda_Q$) as illustrated in Fig. 8c. After the formation of the quasiparticle-mediated charge current via inverse spin Hall effect, the quasiparticles will condensate into Coopers pairs, which gives rise to a charge imbalance effect and a spacing-dependent inverse spin Hall voltage signal. When the voltage probes are far away from the Cu/NbN interface (d2 in Fig. 8c), the charge imbalance or inverse spin Hall signal is not observable. Fig. 8d compares the inverse spin Hall signal for two devices with different distances (d1 = 400 nm and d2 = 10 µm) from the Cu/NbN interfaces measured at $T$ = 3 K with a spin injection current of 1 µA. A tiny inverse spin Hall signal for the device with d2 = 10 µm is observed, which is markedly suppressed compared to the device with d1 = 400 nm.

Both the spin injection current and the voltage probe distance dependences of the inverse spin Hall signal confirm that the inverse spin Hall effect is mediated by the quasiparticles of the superconducting NbN layer. This novel mechanism mediated by quasiparticles gives rise to a huge enhancement of the inverse spin Hall effect, which is more than 2,000 times larger than that in the normal state mediated by electrons in the presence of spin orbit coupling. These results pave the way to realize a sensitive spin detector with superconductors. To use superconductors as spin generators, quasiparticle-mediated spin Hall effect which generates spin current from quasiparticle charge imbalance needs future studies.

**Anomalous Hall and spin Hall effects in non-collinear antiferromagnets**

Since the early work done by Edwin Hall on ferromagnetic metals[125], anomalous Hall effect has been intensively studied, from the mechanisms (intrinsic, skew scattering, side jump scattering) to various materials including ferromagnetic metals, diluted ferromagnetic semiconductors, and heavy fermions, etc[126-128]. Different from the anomalous Hall effect in FM, which is usually assumed to be proportional to its magnetization, the non-collinear antiferromagnets are predicted to hold a large anomalous Hall effect arising from a real-space Berry phase effect despite almost zero net magnetization[128-134]. Chen *et al*[133] performed the calculation on $IrMn_3$, of which Mn sublattices consist of 2D kagome lattices viewed from its (111) direction. They theoretically demonstrate that a large anomalous Hall effect exists, arising from the mirror symmetry breaking and spin-orbit coupling. The total magnetization is calculated to be ~0.02 $\mu_B$ per formula unit along the (111) direction and the intrinsic anomalous Hall conductivity is 218 $\Omega^{-1}cm^{-1}$.

Shortly after the theoretical studies, two experimental groups have observed the large anomalous Hall effect in non-collinear antiferromagnets[135,136]. Nakatsuji *et al*[135] found a large anomalous Hall effect in $Mn_3Sn$, an antiferromagnet that has a 120-degree spin order at room temperature. As shown in Fig. 9a, each *ab* plane consists of a slightly distorted kagome lattice of Mn moments each of ~3 $\mu_B$, and a very small net ferromagnetic moment of ~0.002 $\mu_B$ per Mn atom arises from the geometrical frustration. Since only one of the three Mn moments is parallel to the local easy-axis towards its nearest-neighboring Sn sites, the canting of the other two spins towards the local easy-axis is considered to result in the weak net ferromagnetic moment, which allows the switching of the sign of net magnetization and the anomalous Hall effect via an external magnetic field. By subtracting the ordinary Hall effect, anomalous Hall effect due to magnetization, the anomalous Hall effect due to the non-collinear spin order ($\rho_H^{AF}$) could be obtained, which can be described by the following equation.

$$\rho_H^{AF} = \rho_H - R_0 B - R_s \mu_0 M \qquad (5)$$

where $\rho_H$ is the total Hall resistivity, $R_0$ is the ordinary Hall coefficient, $B$ is the external magnetic field, $R_s$ is the regular anomalous Hall coefficient that is related to magnetization, $\mu_0$ is the permeability, and $M$ is the magnetization. The external magnetic field dependence of $\rho_H^{AF}$ at $T = 300$ K is shown in Fig. 9b, from which a strong facet dependence is observed. When the current is along the crystal's [0001] direction, and the magnetic field is along [01$\bar{1}$0] direction, the largest anomalous Hall effect is observed. Whereas, no anomalous Hall effect is observed when the current is along the crystal's [01$\bar{1}$0] direction, and the magnetic field is along [0001] direction. This facet dependence is consistent with the magnetic field dependence of the magnetization arising from the staggered moments of the non-collinear spin structure. The switching of the magnetization also corresponds to changing the spin-chirality of the non-collinear antiferromagnetic order, which gives rise to the sign switching of $\rho_H^{AF}$. Besides Mn$_3$Sn, a large anomalous Hall effect has also been observed on Heusler compound Mn$_3$Ge[136], which has the similar crystalline structure as Mn$_3$Sn. A much larger anomalous Hall conductivity is observed in Mn$_3$Ge, which is nearly three times higher than that of Mn$_3$Sn. Recently, the anomalous Nernst effect and magneto-optical Kerr effect of Mn$_3$Sn has also been investigated[137-139]. A large transverse Seebeck coefficient ~ 0.35 – ~ 0.6 µVK$^{-1}$ at zero magnetic field is observed from 200 to 300 K. A large zero-field Kerr rotation angle of 20 mdeg is observed at room temperature, comparable to ferromagnetic metals[139]. Also, a large topological Hall effect has been reported in $Mn_5Si_3$, when the spin arrangement changes from collinear to non-collinear as the temperature decreases, suggesting the origin of the topological Hall effect is the non-collinear magnetic structure[140].

The spin Hall effect, the conversion between charge and spin currents, shares the same origin as the anomalous Hall effect: relativistic spin-orbit coupling generates an asymmetric deflection of the charge carriers depending on their spin direction[28,128]. Both recent experimental and theoretical studies have demonstrated the considerably large spin Hall effect due to the non-collinear spin order[141,142]. As shown in Fig. 9c, the spin-up and spin-down carriers flow in the opposite directions, resulting in the generation of a spin current. Zhang et al[141] studied the charge to spin conversion in single crystalline $IrMn_3$ films by measuring the spin orbit torque on a thin Py layer at room temperature using the ST-FMR technique. $IrMn_3$ is known to display a triangular chiral magnetic structure with the Mn magnetic moments aligned at 120° to each other in the (111) plane when the Ir and Mn atoms are chemically ordered. In the cubic $IrMn_3$ lattice the Mn atoms are arranged in the form of triangles within the (111) plane of the primitive unit cell. A large spin Hall angle is observed in (001) oriented single crystalline antiferromagnetic $IrMn_3$ thin films, which is ~ two times larger than the (111) oriented $IrMn_3$ films. The strong facet dependence of the effective spin Hall angle of $IrMn_3$ (Fig. 9d) identifies two distinct mechanisms for generation of spin current from charge carriers; the first mechanism, which is facet independent, arises from conventional bulk spin-dependent scattering within the $IrMn_3$ layer. And the second intrinsic mechanism derives from the non-collinear antiferromagnetic structure. The effective spin Hall effect of the second mechanism can be dramatically enhanced by manipulating the populations of the various antiferromagnetic domains, ie, a ~ 75% enhancement of the effective spin Hall angle is achieved by perpendicular magnetic field annealing on the 3-nm (001) IrMn$_3$ thin film. The reconfiguration of the antiferromagnetic domain structure of the IrMn$_3$ layer can be reconfigured due to exchange coupling with the Py layer when it is cooled from a temperature above its blocking temperature to room temperature. Further experimental evidences also support the novel mechanism of the spin Hall effect due to the non-collinear spin order. The insertion of a thin Au layer is used to weaken the exchange coupling between Py and IrMn$_3$, resulting in the largely reduced dependence of the effective spin Hall angle on the crystalline orientations. Besides, the facet dependence of the effective spin Hall angle disappears for CuAu-I-type antiferromagnets (Ir$_{1-x}$Mn$_x$ with x ~ 53% and 64%) with collinear spin order.

One of the possible origins of these interesting spin-dependent properties is associated to significant enhancement of the Berry curvature, particularly around band crossing points called Weyl points near the Fermi energy[135,137,141-143]. Direct experimental evidences of the Mn$_3$Sn and Mn$_3$Ge as Weyl semimetals, such as surface Fermi Arcs via spin-APRES, have not been reported yet[143]. The proposal using the spin orbit torque and antiferromagnetic order to control the topological properties of Weyl semimetals is quite intriguing[144].

## Summary and outlook

In conclusion, the conversion between spin and charge using quantum materials has attracted a great deal of attention recently and quantum materials have exhibited unique spin dependent properties. Looking forward, novel mechanisms that can generate spin current would be particularly interesting, arising from crystalline symmetry, or coupling with other degrees of freedom, such as phonon, magnon, spinon, and valley/layer degree of freedom, etc[103,145-148]. The spin and charge conversion in these quantum materials could be also very important for the detection of the spin current generated via various mechanisms[146], the probe of the non-trivial magnetic fluctuations[149-152], and search for new quantum states[153,154]. Besides, the efficient spin current generation may pave the way for future spintronics device applications, including magnetic random accessory memory, domain wall Racetrack memory, and skyrmion Racetrack memory, etc[155-158].

**Acknowledgements**

We are grateful to Wenyu Xing and Yunyan Yao for their help preparing illustrations. W.H. acknowledges the financial support from National Basic Research Programs of China (Grant Nos. 2015CB921104 and 2014CB920902), the 1000 Talents Program for Young Scientists of China, and National Natural Science Foundation of China (No. 11574006). Y.O. acknowledges the financial support from Grant-in-Aid for Scientific Research on Innovative Area, "Nano Spin Conversion Science" (Grant No. 26103002) and CREST (Grant No. JPMJCR15Q5). S.M. acknowledges the financial support from ERATO-JST (JPMJER1402), and KAKENHI (No. 26103006 and No. JP17H02927) from MEXT, Japan.

**Author Contributions**

The manuscript was written through contributions of all authors. W.H. wrote the paper and Y.O. and S.M. examined and improved it.

**Competing interests**

The authors declare no competing both financial and non-financial interests.

**Figure Legends**

**Figure 1. Pure spin current. a**, Illustration of pure spin current. Spin-up and spin-down elections move in the opposite directions, resulting in the flow of pure spin current without companying any charge current. **b**, Pure spin current in the spin transistors or logic devices. Two ferromagnetic or heavy metallic electrodes are used for pure spin current injection/detection. The pure spin current could be injected via electrical charge current spin injection, spin pumping, spin Seebeck effect, and spin Hall effect, etc. Pure spin current flows in the spin channel that carries information. **c**, Illustration of the diffusion of pure spin current in spin channels with the presence of spin scattering. $\mu_s$ is the spin dependent chemical potential. **d**, Schematic of spin orbit torque at the interface between a non-magnetic metal (NM) with large spin orbit coupling and a ferromagnet (FM). **e**, Illustration of the spin transfer torque from spin current on FM.

**Figure 2. Quantum materials for spin and charge conversion.** A schematic illustration of the spin and charge conversion in emergent quantum materials, including Rashba interfaces, topological insulator surface states, two-dimensional materials, and quasiparticles in superconductors, as well as antiferromagnets with non-collinear spin order.

**Figure 3. Edelstein effect and inverse Edelstein effect of Rashba interfaces. a**, The energy dispersion of Rashba interfaces. At the Fermi level, the outer and inner Fermi circles exhibit opposite spin textures; one is clockwise, and the other one is couterclockwise. **b, c**, Illustration of Edelstein effect (conversion from charge to spin) and inverse Edelstein effect (conversion from spin to charge). **d**, Measurement setup of Edelstein effect at the Bi/Ag interface via spin-polarized positron beam. **e**, The surface spin polarization as a function of the Bi thickness in the Bi/Ag (25 nm)/Al$_2$O$_3$ samples. Solid line represents an exponential decay. **d**, Measurement setup

of inverse Edelstein effect at the Bi/Ag interface. The spin current is generated via spin pumping from NiFe, and then converted to charge current. **e**, The magnetic field dependence of the charge current measured on Ag (10 nm)/NiFe (15 nm), Bi (8 nm)/NiFe (15 nm) and Bi (8 nm)/Ag (5 nm)/NiFe (15 nm) samples, respectively. Red lines are Lorentzian fits. (**d, e** adapted from ref. 34 with permission, copyright American Physical Society 2015) (**f, g**, adapted from ref. 33 with permission, copyright Springer Nature 2013)

**Figure 4. Inverse Edelstein effect in Rashba states at oxides interface and metal/oxide interface. a-b,** Illustration of Rashba-split two dimensional electron gas (2DEG) between two insulating oxides of $LaAlO_3$ and $SrTiO_3$. **c**, Room temperature gate tuning of inverse Edelstein effect for the Rashba-split 2DEG between $SrTiO_3$ and 3-unit cell $LaAlO_3$. Inset: Schematic of the inverse Edelstein effect measurement set up where Py is the spin pumping source and the gate voltage is applied across the insulating $SrTiO_3$ substrate. **d**, Gate dependence of spin and charge conversion efficiency ($\lambda_{IEE}$) of the $SrTiO_3$/2-unit cell $LaAlO_3$/Py sample at $T = 7$ K. Inset : sketch of the multiband structure for the Rashba 2DEG at the oxide interface **e**, Measurement setup of inverse Edelstein effect at the $Cu/Bi_2O_3$ interface via spin pumping. **f**, Magnetic field dependence of voltage converted from the spin current measured on the Py (5 nm)/Cu (10 nm)/$Bi_2O_3$ (100 nm) sample (red curve) and a control Py/Cu sample (black curve). Solid line represents Lorentian fit of the raw data. (**b, c,** adapted from ref. 38 with permission, copyright American Association for the Advancement of Science 2017) (**d**, adapted from ref. 37 with permission, copyright Springer Nature 2016) (**e, f,** adapted from ref. 42 with permission, copyright Japan Society of Applied Physics 2016)

**Figure 5. Spin and charge conversion in topological insulator surface states. a**, Schematic of spin-momentum locking in the Dirac bands of topological insulator surface states. **b**, **c**, (Top)

Spin accumulation due to Edelstein effect of the spin-momentum locked topological surface states for n-type (**b**) and p-type (**c**) $(Bi_{1-x}Sb_x)_2Te_3$. Solid and dashed circles are the Fermi circles with and without electric fields, respectively. (Bottom) Difference in the Fermi distribution on applying an electric field. **d,** Spin current generation and detection via spin-torque ferromagnetic resonance (ST-FMR) in $(Bi_{1-x}Sb_x)_2Te_3$ (8 nm)/Cu (8 nm)/Py (10 nm) trilayer structure. **e,** Dependence of the charge to spin conversion efficiency of $(Bi_{1-x}Sb_x)_2Te_3$ on the Sb composition (*x*). Insets: Fermi level position for each Sb doped $(Bi_{1-x}Sb_x)_2Te_3$ sample. (**b-e,** adapted from ref. 72 with permission, copyright Springer Nature 2016)

**Figure 6. Spin and charge conversion in new types of 3D topological insulators. a**, Surface states of α-Sn (001) films (~ 30 monolayers) measured by ARPES. α-Sn (001) films are grown on InSb (001) substrate by molecular beam epitaxy. **b**, The FMR spectrum and charge current measured on InSb/α-Sn/Ag/Fe/Au sample. Ag layer is used to protect the topological surface states from degradation by the growth of Fe layer. **c**, Comparison of topological Kondo insulator ($SmB_6$) with $Bi_2Se_3$ type topological insulators. Inset: crystalline structure of $SmB_6$. TSS/BCB stands for topological surface states and bulk conducting bands. **d**, The resistance of the $SmB_6$ single crystal as a function of temperature. Green background highlights the region where pure surface states exist. Inset: measurement of the spin and charge conversion via spin pumping from Py. **e**, Magnetic field dependence of voltages on the $SmB_6$/Py device measured at *T* = 0.84, 1.66, 2.1, 2.3, and 10 K, respectively. The measurement is performed under microwave frequency of 10.1 GHz. (**a, b,** adapted from ref. 87 with permission, copyright American Physical Society 2016) (**c,** adapted from ref. 93 with permission, copyright Springer Nature 2014) (**d, e,** adapted from ref. 96 with permission, copyright Springer Nature 2016)

**Figure 7. Spin and charge conversion in 2D materials. a,** Spin to charge conversion in graphene on yttrium iron garnet (YIG), a ferrimagnetic insulator. The spin current is generated from spin pumping from YIG, and converted to charge current in graphene. **b,** Magnetic field dependence of the spin pumping voltage measured on YIG/Graphene. **c,** Current induced out-of-plane spin-orbit torque in $WTe_2$. The current is applied along the a-axis of a $WTe_2$/Py bilayer. **d,** Antisymmetric voltages from ST-FMR measurements for a $WTe_2$ (5.5 nm)/Py (6 nm) device as a function of in-plane magnetic field angle. The microwave frequency is 9 GHz and the applied microwave power is 5 dBm. (**a, b,** adapted from ref. 105 with permission, copyright American Physical Society 2015) (**c, d,** adapted from ref. 111 with permission, copyright Springer Nature 2016)

**Figure 8. Quasiparticle-mediated large spin Hall effect in a superconductor. a,** Schematic of spin Hall measurement on a Cu nonlocal spin valve device. The pure spin current ($J_S$) is absorbed at NbN/Cu interface and then converted to quasiparticle charge current ($J_Q$) via inverse spin Hall effect. **b,** Comparison of current dependence of quasiparticle-mediated spin Hall effect at $T = 3$ K and normal spin Hall effect at $T = 20$ K. **c,** Spatial decay profile of the voltage generated by the inverse spin Hall effect. $\lambda_Q$ is the quasiparticle charge imbalance length, and d1 and d2 are distances between the voltage probes and the NbN/Cu junction. **d,** Inverse spin Hall resistance as a function of magnetic field from the samples with distance of d1=400 nm (red) and d2=10μm (green) at $T = 3$ K. (**a-d,** adapted from ref. 122, , with permission, copyright Springer Nature 2015)

**Figure 9. Spin and charge conversion in non-collinear antiferromagnets. a,** Crystalline and magnetic structures of $Mn_3Sn$ in the *ab* plane. **b,** Large anomalous Hall effect in $Mn_3Sn$ arising from the non-collinear antiferromagnetic order ($\rho_H^{AF}$) at room temperature. **c,** Illustration of the spin Hall effect arising from the non-collinear 120-degree spin order. **d,** Facet-dependent large spin Hall effect in antiferromagnet IrMn$_3$. Olive, blue, and cyan circles indicate (001)-oriented, (111)-oriented, and polycrystalline IrMn$_3$ thin films. The effective spin Hall angles are obtained by measuring the spin orbit torque in IrMn$_3$/Py bilayer devices via ST-FMR technique. **(a, b,** adapted from ref. 135, with permission, copyright Springer Nature 2015) **(c, d,** adapted from ref. 141 with permission, copyright American Association for the Advancement of Science 2016)

Figure 1

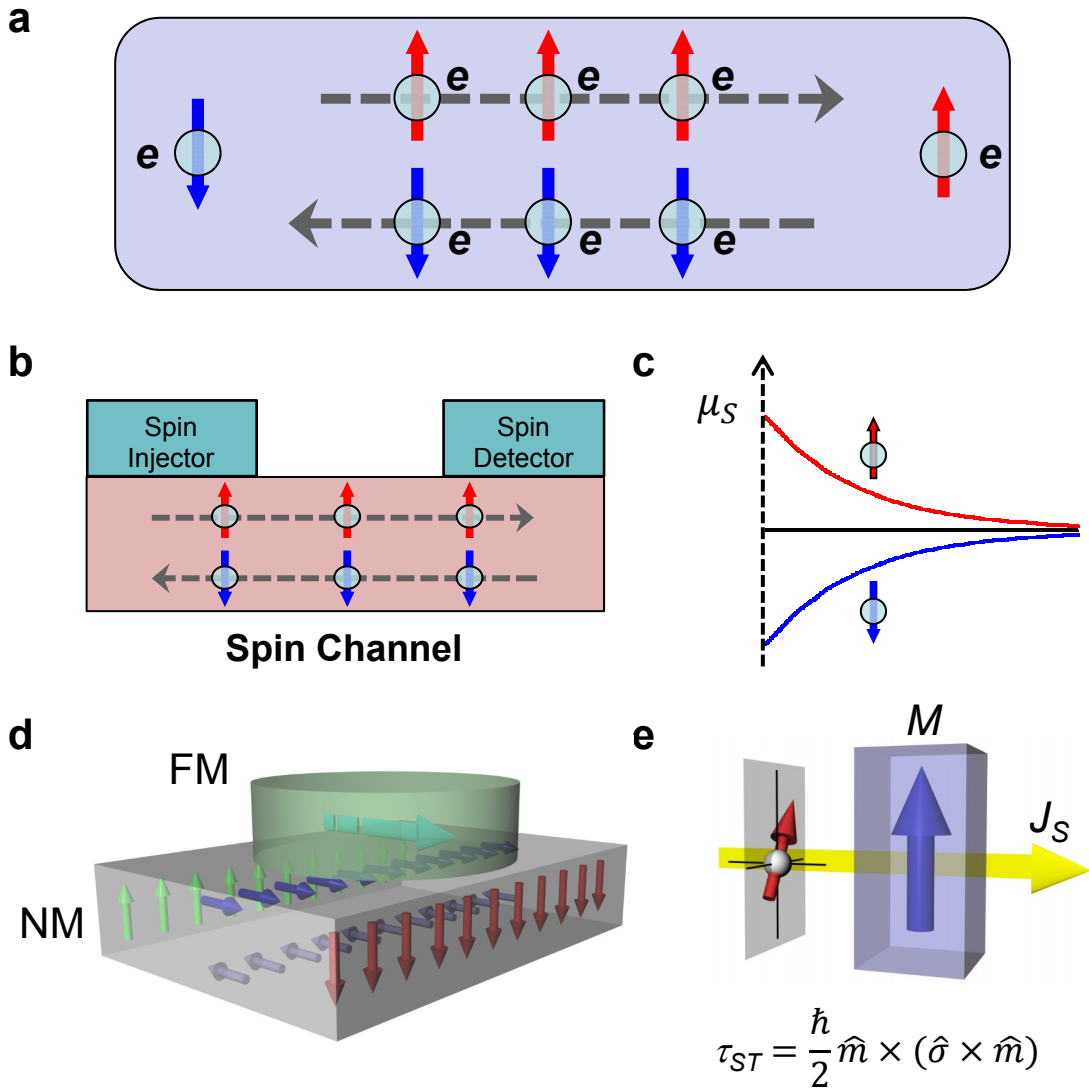

Figure 2

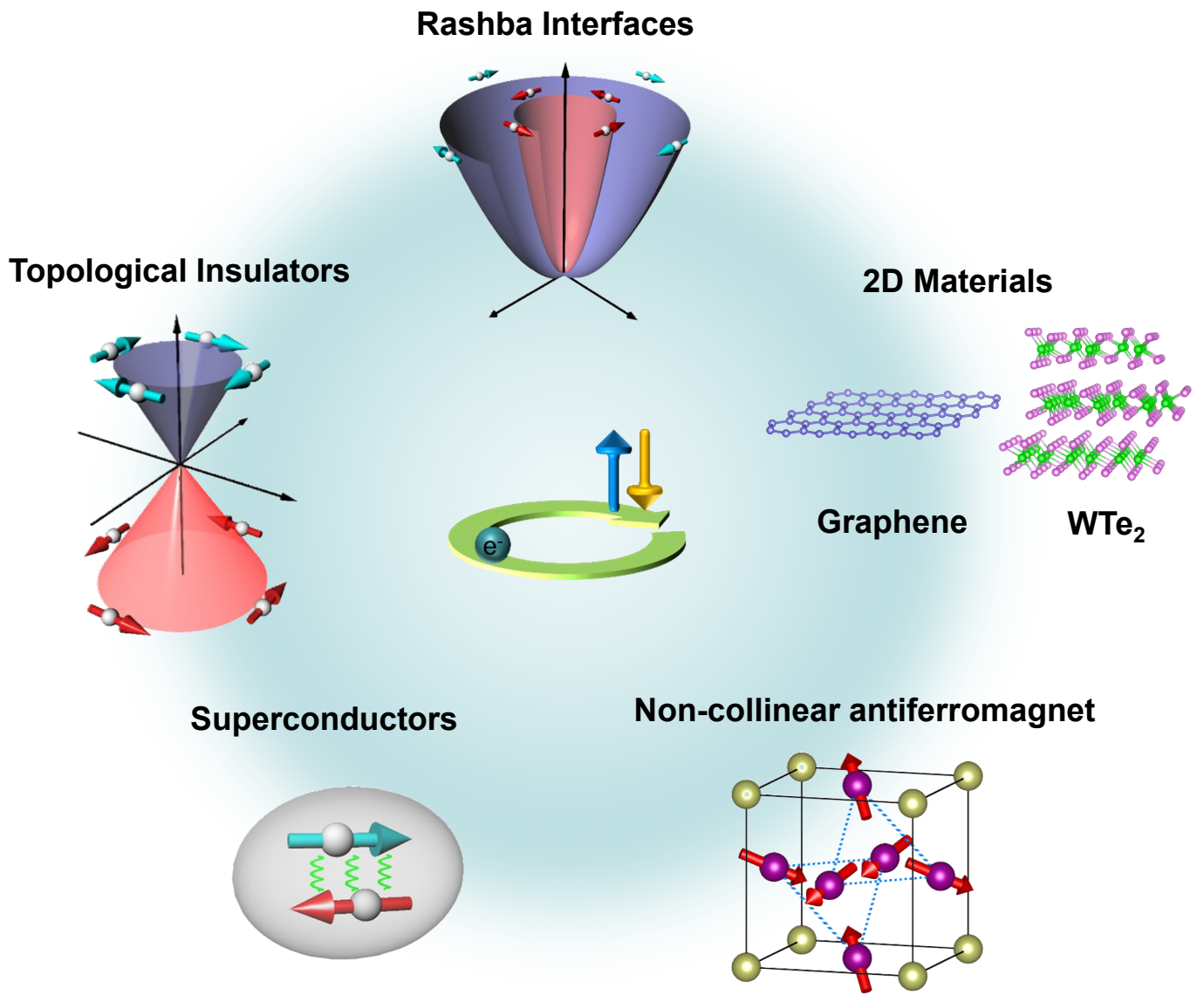

Figure 3

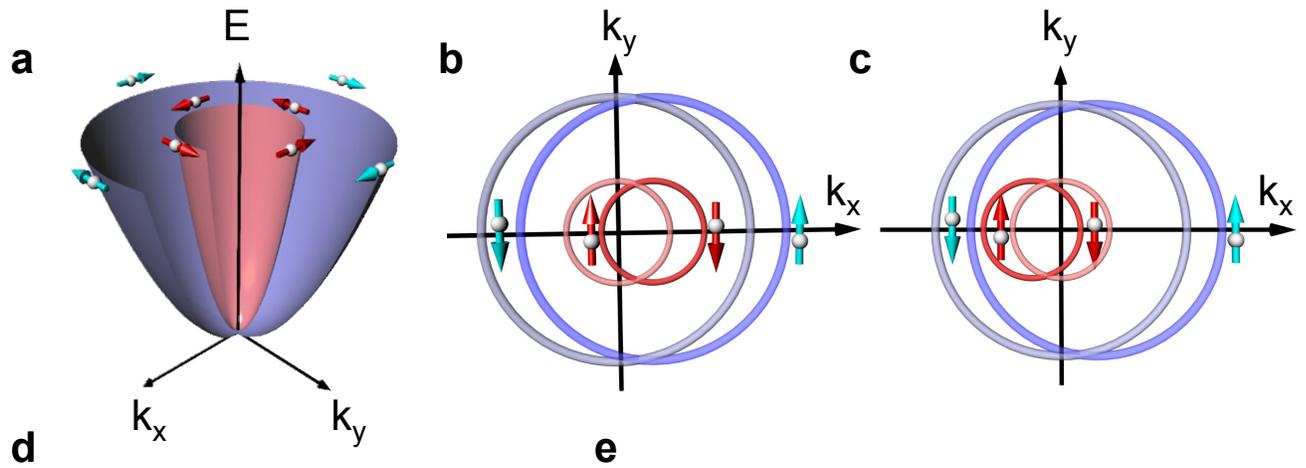

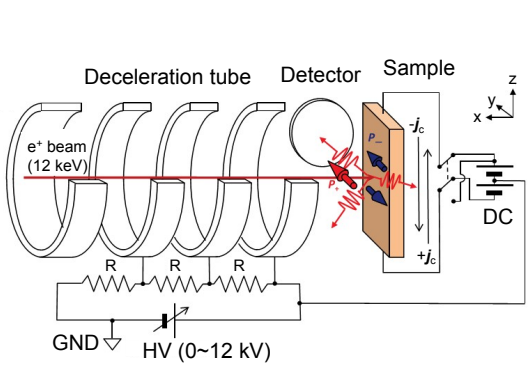

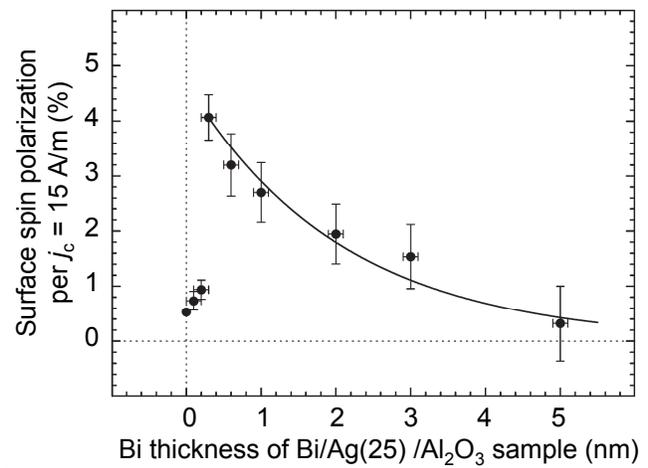

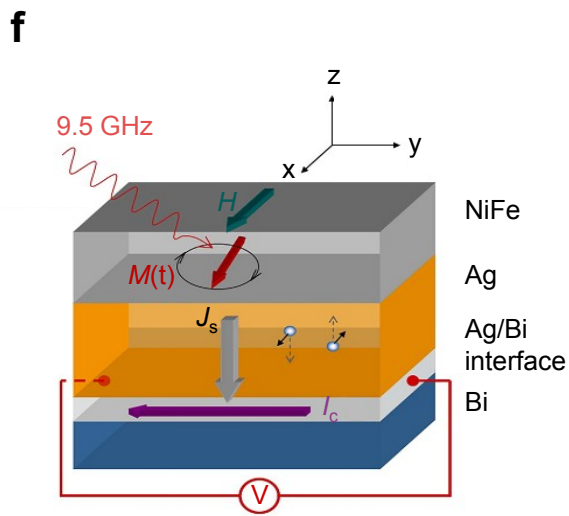

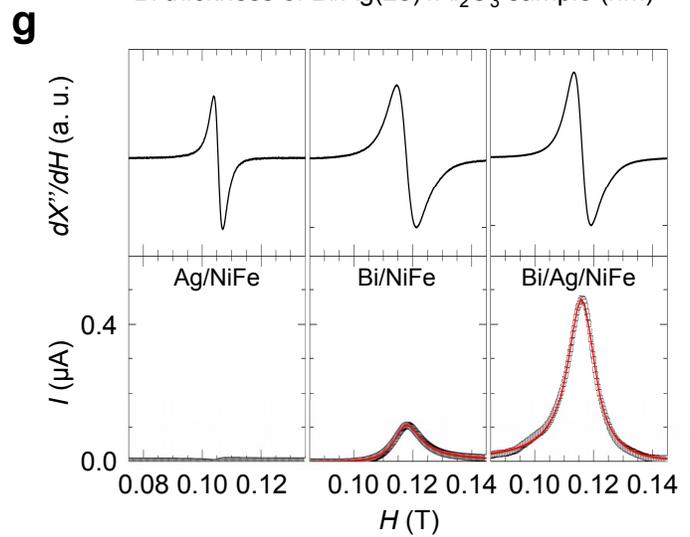

Figure 4

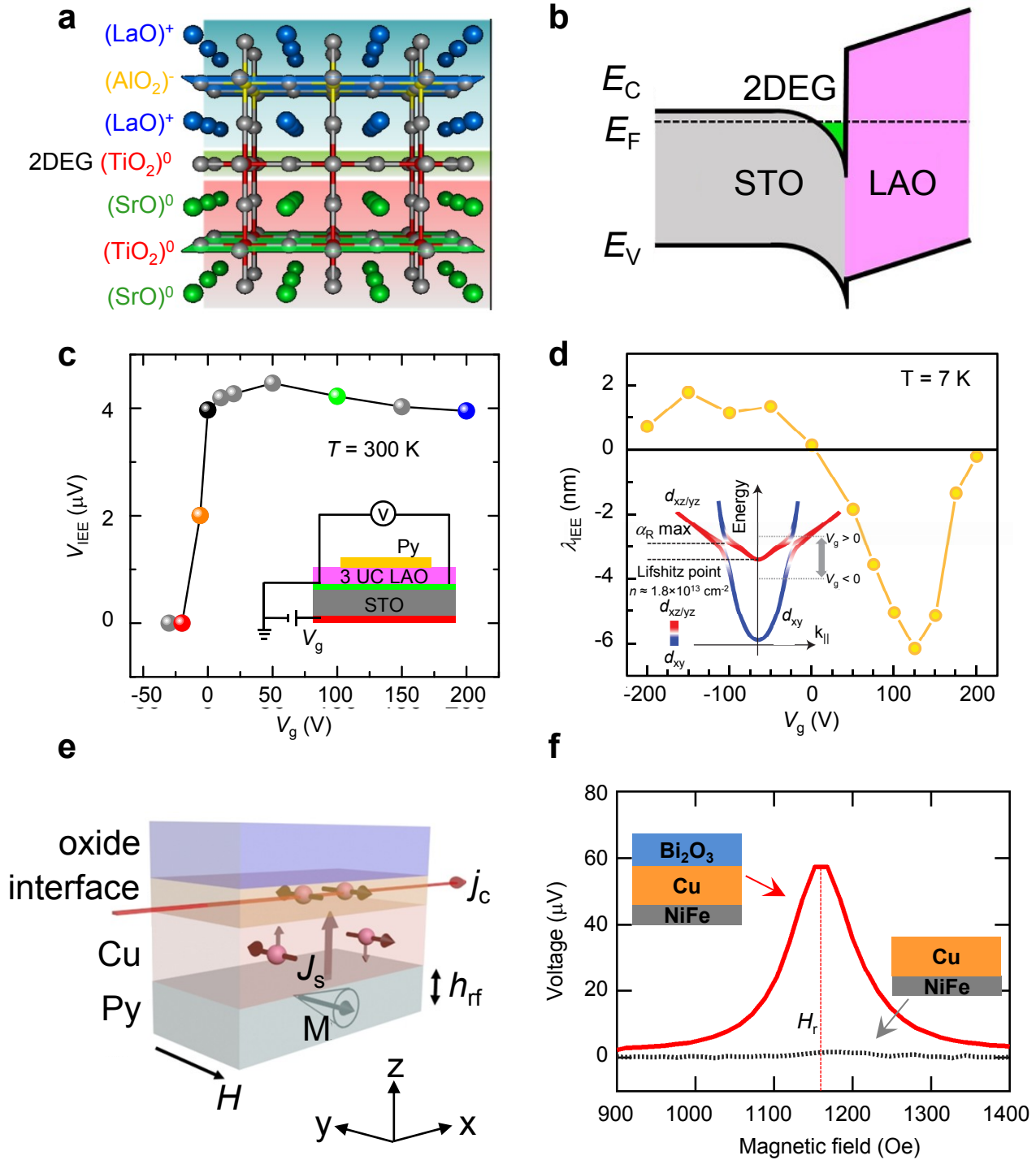

Figure 5

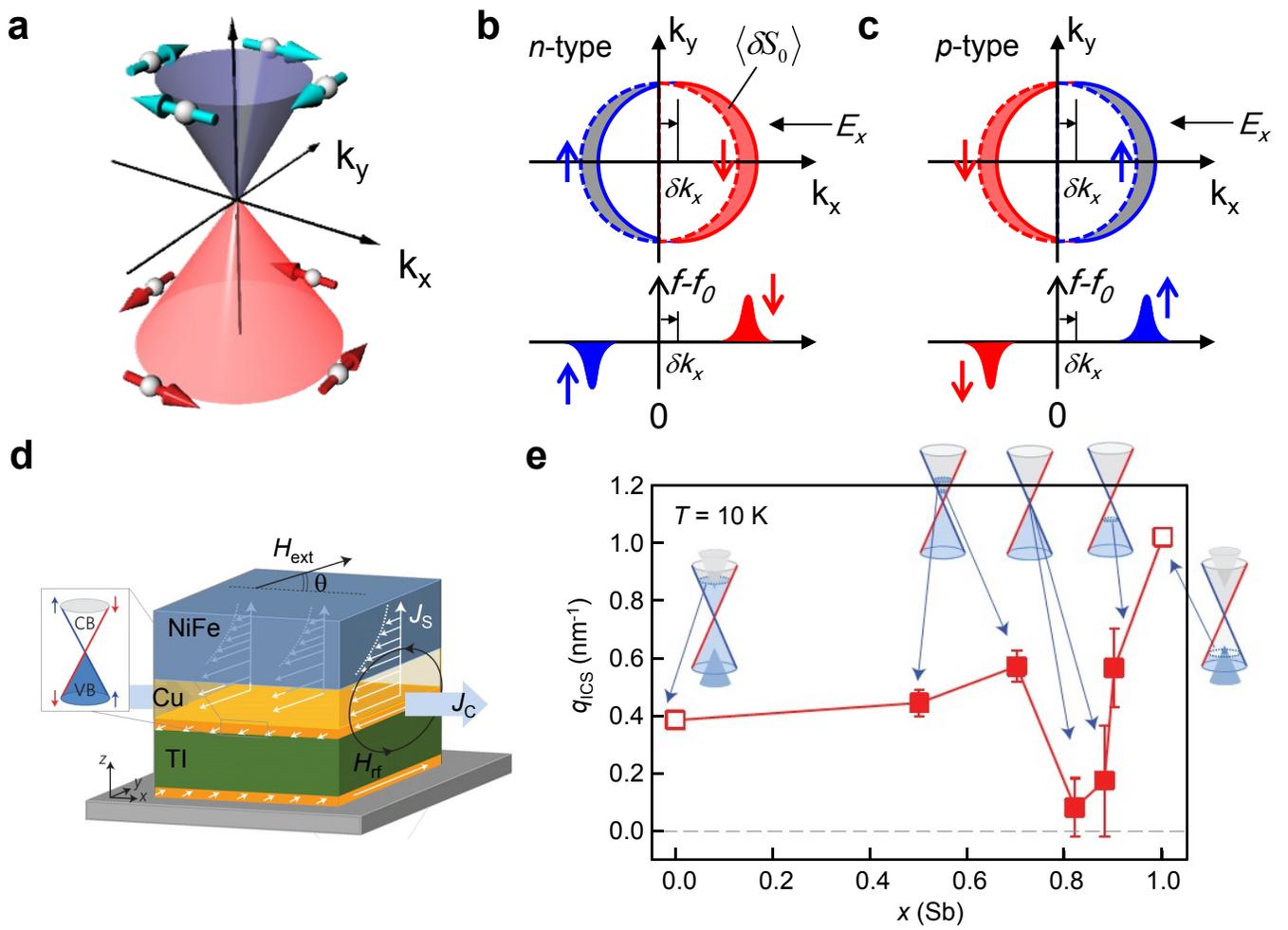

Figure 6

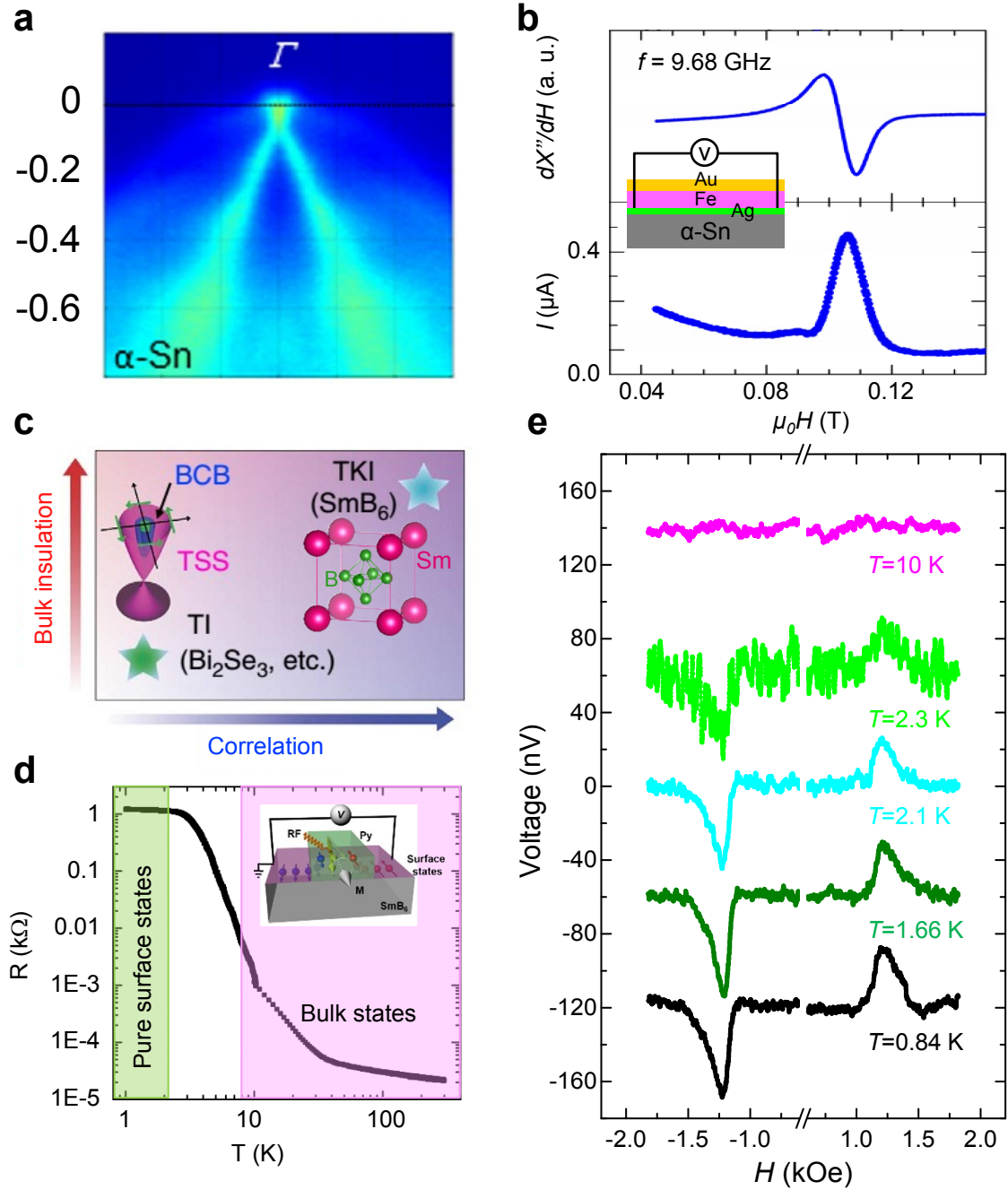

Figure 7

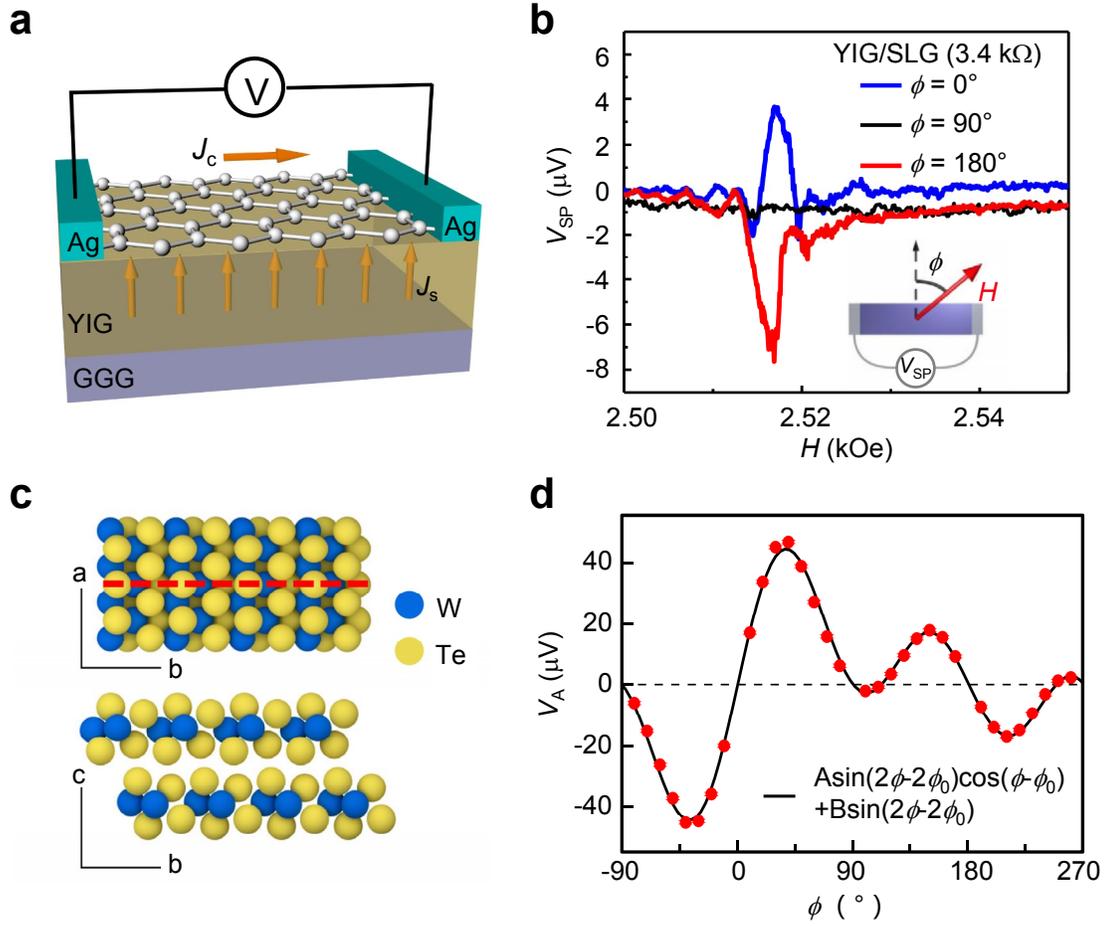

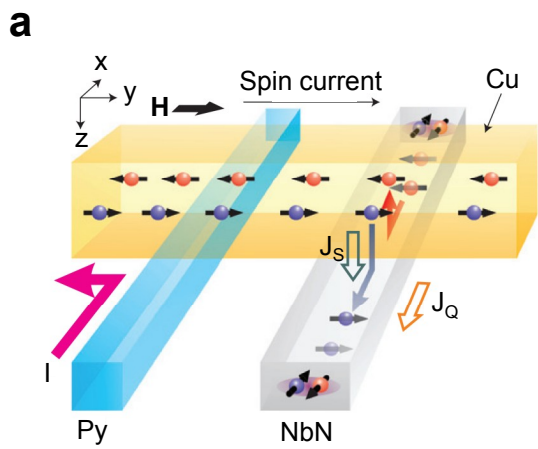
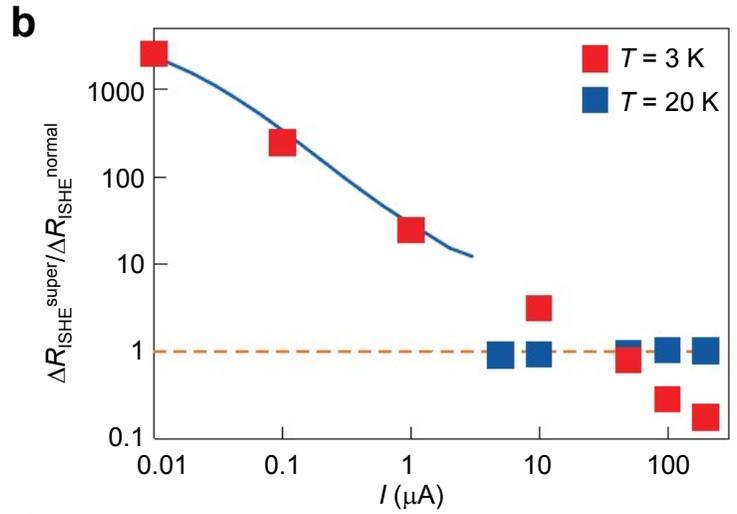
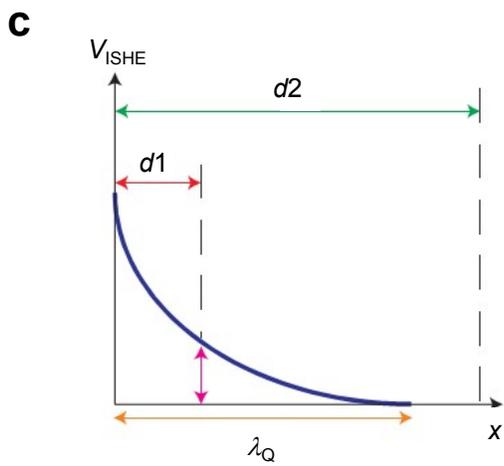
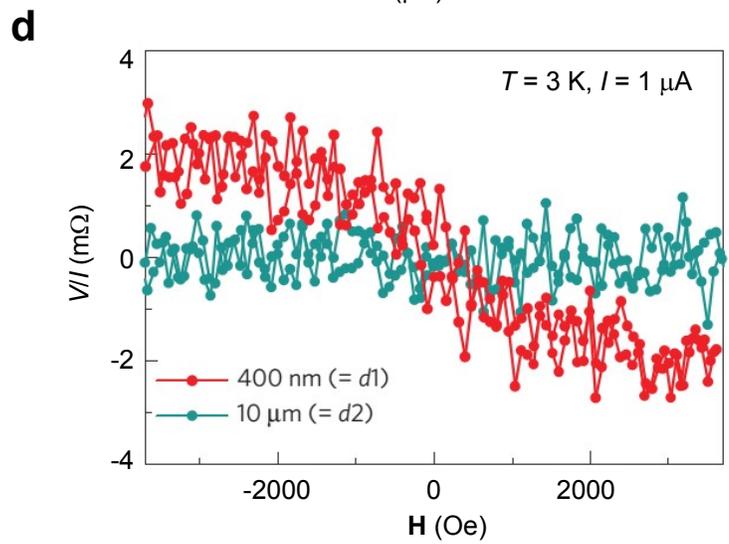

Figure 9

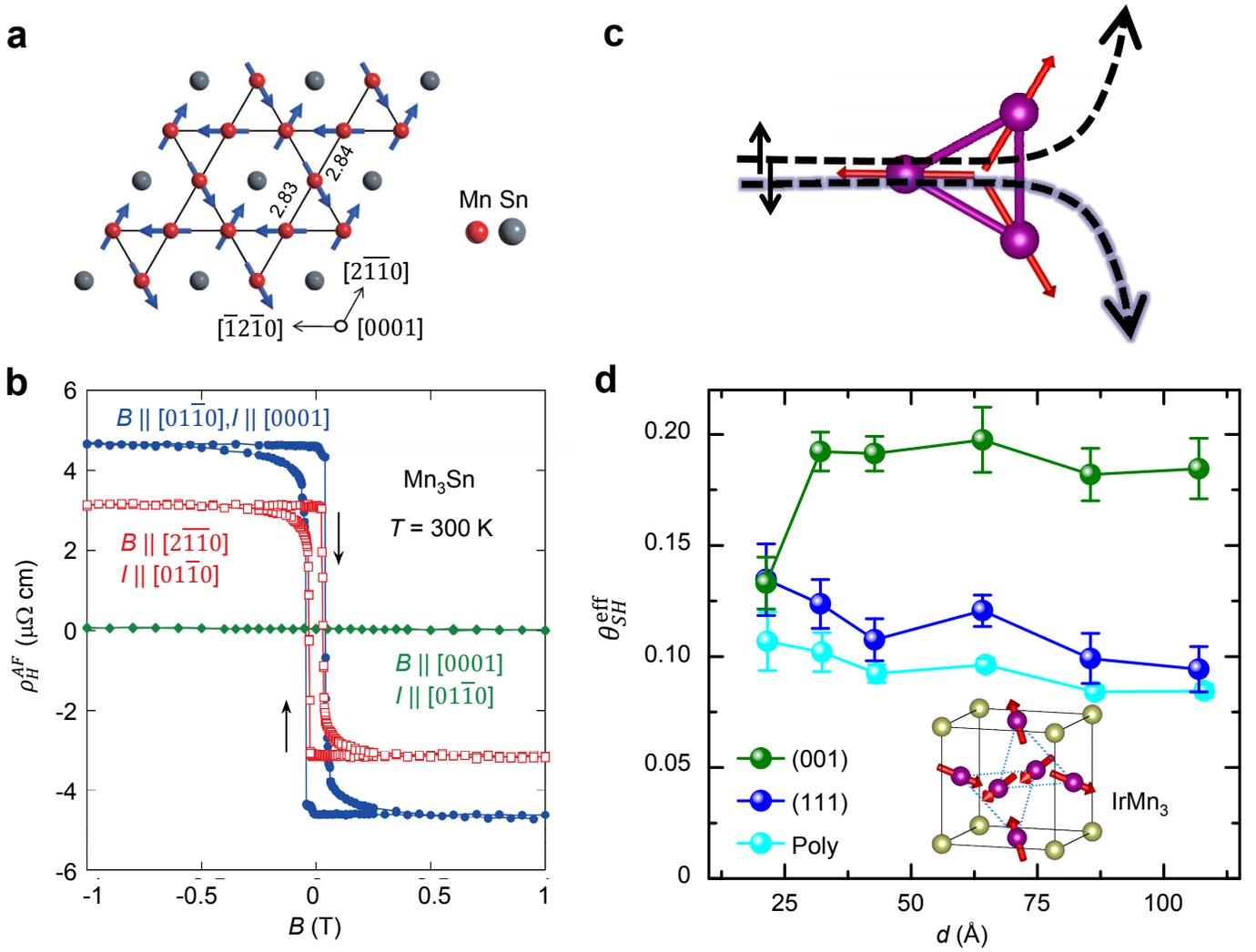